\documentclass[letterpaper, 10 pt, conference]{ieeeconf}
\IEEEoverridecommandlockouts

\usepackage{amsmath}
\usepackage{amssymb}
\usepackage{mathrsfs}
\usepackage{graphicx}
\usepackage{comment}
\usepackage{booktabs}

\title{
    \LARGE \bf
    Inertia Partitioning Modular Robust Control Framework for Reconfigurable Multibody Systems
}

\author{
    Mohammad Dastranj$^{1,*}$ and Jouni Mattila$^{1}$%
    \thanks{
        \protect\rule{0.965\linewidth}{0.4pt} \indent
        $^{1}$The authors are with the Unit of Automation Technology and Mechanical Engineering,
        Faculty of Engineering and Natural Sciences, Tampere University, 33720 Tampere, Finland
        {\tt\small \{mohammad.dastranj, jouni.mattila\}@tuni.fi}
    }%
}

\begin{document}

\maketitle

\begin{abstract}
A novel modular modeling and control framework based on Lagrangian mechanics is proposed for multibody systems, motivated by the challenges of modular control of systems with closed kinematic chains and by the need for a modeling framework that remains locally updatable under reconfiguration of body-level geometric and inertial properties. In the framework, modularity is defined with respect to the degrees of freedom of the multibody system, represented in the model by the minimal generalized coordinates, and the inertial properties of each body are partitioned with respect to how they are reflected in the kinetic energy of the system through the motion induced by each degree of freedom. By expressing body contributions through body-fixed-frame Jacobians and spatial inertia matrices, the dynamic model remains locally updatable under changes in geometric and inertial parameters, which is advantageous for reconfigurable multibody systems. For multibody systems in which a mapping between the auxiliary and minimal generalized coordinates is available, the approach accommodates closed kinematic chains in a minimal-coordinate ordinary-differential-equation form without explicit constraint-force calculation or differential-algebraic-equation formulation. Based on the resulting modular equations of motion, a robust model-based controller is designed for trajectory tracking, and practical boundedness of the tracking error is analyzed under bounded uncertainty and external disturbance. The proposed framework is implemented in simulation on a three-degree-of-freedom series-parallel manipulator, where uncertainties and disturbances are introduced to assess robustness. The results are consistent with the expected stability and tracking performance, indicating the potential of the framework for trajectory-tracking control of reconfigurable multibody systems with closed kinematic chains.
\end{abstract}


\section{Introduction}

Advances in the design of multibody dynamic systems have led to greater complexity in their structures, which directly affects their modeling and control. With more complex multibody systems, a modular and systematic framework is noticeably helpful as it supports a fixed procedure toward modeling and control of such systems. This fact is remarkably beneficial when a multibody system can be reconfigured \cite{Cao2007ReconfigurableControlStructure}, i.e. the geometric and inertial parameters of one or multiple bodies in the system can be altered. As an example, this is a prospective scenario in upgrading existing multibody systems. Therefore, the motivation arises to adopt a modular framework to control multibody systems. In the present work, reconfiguration is addressed in the sense that changes in body-level geometric or inertial properties can be incorporated through local updates of the corresponding Jacobians and spatial inertia terms, followed by reassembly of the global dynamics without re-deriving the full model from the beginning.

\subsection{Background}

There are well-established modular control frameworks for multibody systems in the literature, notably passivity-based control of port-Hamiltonian systems \cite{vanDerSchaft2017} and virtual decomposition control (VDC) \cite{zhu2010virtual}. In passivity-based control of port-Hamiltonian systems, modularity commonly refers to a system being represented as interconnected subsystems, each equipped with its own energy-based or port-Hamiltonian model and interacting through power-preserving ports or interconnections \cite{vanDerSchaft2017,berger2025port,brugnoli2021port}. In contrast, VDC interprets modularity through a virtual decomposition of the robotic system into subsystems, such as bodies and joints, whose dynamic interactions are represented across virtual cutting points by forces and moments exchanged between adjacent subsystems \cite{zhu2010virtual}.

The port-Hamiltonian framework provides a rigorous energy-based representation of physical systems, where passivity follows from the system formulation and where interconnections are defined in a power-preserving manner \cite{vanDerSchaft2017}. Within this framework, which has been applied to classes of rigid \cite{berger2025port} and flexible \cite{brugnoli2021port} multibody systems, passivity properties are used for stabilization and controller design  \cite{ortega2002interconnection}.

Alternatively, VDC decomposes a multibody system into simpler subsystems by introducing virtual cutting points. The subsystem interconnections are handled through virtual power flows, and local controllers are designed at the subsystem level. Stability of the overall system is then analyzed through the notion of virtual stability, whereby the virtual power flows associated with adjacent subsystems cancel across the cutting points \cite{zhu2010virtual}.

\subsection{Literature Review}

Advances in heavy-duty robotic systems, such as those in construction, have introduced architectures with extended reach and multi-ton payload capacities \cite{KOIVUMAKI201559}. These systems often employ hybrid series–parallel structures driven by hydraulic or electro-mechanical actuators. However, the inclusion of closed kinematic chains in such mechanisms introduces additional challenges to model-based modular control.

Regarding the port-Hamiltonian framework, this issue is addressed in \cite{vanderschaft2018generalizedporthamiltoniandaesystems} and \cite{van2020dirac}, and the result is a system of differential-algebraic equations (DAEs). The addition of constraint equations to form the system of DAEs imposes numerical challenges and stability issues in forward dynamics \cite{fernandez2022non}, including constraint drift and the need for constraint stabilization \cite{khoshnazar2024application}. Furthermore, DAE-based formulations can complicate model-based control design, as the resulting inverse models may have high differentiation index and thus require careful numerical treatment \cite{drucker2023trajectory}.

VDC also relies on modifications to its general framework to handle closed kinematic chains, as it was originally developed for multibody systems with tree-like structure by using the spatial-vector form of the Newton-Euler formalism \cite{zhu2010virtual}. Currently, no widely adopted solution exists within the VDC framework for calculating the constraint forces and torques in closed kinematic chains, and the available modifications are often system-specific, depending on the particular structure of the closed kinematic chain. For instance, \cite{petrovic2022mathematical,ding2023high,barjini2025surrogate}, and \cite{zhang2025equivalence} have provided solutions for utilizing VDC for a certain class of multibody systems including triangular closed kinematic chains comprised of several passive revolute joints and one hydraulically- or electrically-actuated prismatic joint, which are common in heavy-duty robotic manipulators.

These frameworks, despite the solid foundations they provide for modular control, face challenges with closed kinematic chains. These challenges motivate the development of a unified modular control framework for rigid multibody systems with closed kinematic chains which, provided that the mapping between the auxiliary and minimal generalized coordinates is available, avoids the computational burdens commonly associated with DAE formulations.

\subsection{Contributions}
While the term module in the related literature typically refers to a physical component or a small set of physical components, the term is used differently in the present work. In this paper, the modular viewpoint is defined with respect to the degrees of freedom of the multibody system. More precisely, each module corresponds to one degree of freedom together with the system motion induced by that degree of freedom, represented mathematically through the corresponding minimal generalized coordinate. The contribution of each rigid body to such a module is not introduced by physically detaching the body from the system, but by expressing, through body-fixed-frame Jacobians, how that body's inertia is reflected in the kinetic energy associated with the coordinate-induced motion. Accordingly, the proposed modularity is motion-based rather than relying on body detachment. In this sense, the degree of freedom is the conceptual module, while the corresponding minimal generalized coordinate provides its mathematical representation in the dynamic model.

The Jacobian matrices of body-fixed frames of reference are central to this framework. They are necessary to calculate the generalized inertia matrices, which capture how the inertial properties of each body contribute to the kinetic energy in the minimal coordinates, encapsulating the interconnection between system components. Also, the Jacobians are used to calculate the generalized forces associated with external wrenches exerted on the system. Modularity in this sense provides a systematic route for assembling the dynamic model from body-level contributions and thereby facilitates controller synthesis for multi-input systems. At the same time, the use of minimal coordinates yields a system of ordinary differential equations (ODEs) instead of DAEs. Accordingly, the novelty of the proposed approach lies not in constructing independent local controllers for detached subsystems, but in obtaining a globally usable dynamic model through modular assembly of body-level inertial contributions in the minimal-coordinate representation, which is particularly advantageous for closed-chain and reconfigurable multibody systems. The inertia partitioning framework alleviates the computational issues inherent to DAEs as well as the challenges in using the DAE form in model-based control and stability analysis.

In addition, the framework is suitable for reconfigurable multibody systems in the sense that local changes in body geometry or inertial parameters are reflected through local Jacobian and spatial-inertia updates, which can then be reassembled into the global dynamics without reconstructing the entire model.

Based on the modularly assembled equations of motion, a robust model-based controller is designed for trajectory tracking of multibody systems. The controller combines the nominal model-based term with linear error feedback and a robust correction term, and a practical stability analysis is provided under bounded uncertainty and external disturbance. The proposed approach is therefore not limited to modular dynamic modeling, but also provides a systematic path from the modular model to robust controller design for multibody systems with closed kinematic chains, provided that the mapping between the auxiliary and minimal generalized coordinates is available.

\subsection{Paper Structure}
The remainder of this paper is structured as follows. Section \ref{sec:dyn} explains the derivation of the Jacobians, the generalized inertia matrices, and the equations of motion. Section \ref{sec:control} presents the controller design procedure based on the modular dynamic model of the multibody system and provides the corresponding practical stability analysis under bounded uncertainty. Section \ref{sec:simulation} models and controls a three-degree-of-freedom (3-DoF) series-parallel manipulator in simulation by implementing the inertia partitioning framework, and evaluates the trajectory-tracking performance in the presence of uncertainties and disturbances. Finally, Section \ref{sec:conclusion} concludes the steps taken in the paper and the results that they yield with respect to the intended goals of the research.


\section{Modular Dynamic Modeling of Multibody Systems}
\label{sec:dyn}

In the first part of this section, kinematic analysis is performed using the spatial vector methodology in \cite{lynch2017modern}, whose results are required in the next part, where it is explained how the local generalized inertia matrix is obtained for each rigid body by using the spatial Jacobian corresponding to the motion of an arbitrary body-fixed frame of reference on the same body. Later, the equations of motion for each degree of freedom are obtained.

\subsection{Kinematics of the Multibody System}

For two frames of reference $\{\mathcal{A}\}$ and $\{\mathcal{B}\}$, the homogeneous transformation matrix comprised of the rotation matrix $\boldsymbol{R}_{\mathcal{A}\mathcal{B}}$ and the distance vector $\boldsymbol{p}_{\mathcal{A}\mathcal{B}}$ of frame $\{\mathcal{B}\}$ with respect to frame $\{\mathcal{A}\}$ and expressed in frame $\{\mathcal{B}\}$ is formed as
\begin{equation*}
    \boldsymbol{\mathcal{T}}_{\mathcal{A}\mathcal{B}} =
    \begin{bmatrix}
        \boldsymbol{R}_{\mathcal{A}\mathcal{B}} & \boldsymbol{p}_{\mathcal{A}\mathcal{B}} \\
        \boldsymbol{0} & 1
    \end{bmatrix}
    \in SE(3)
    \label{eq,Tmat}
\end{equation*}
where $SE(3)$ is the Special Euclidean group. In case of $f$ intermediate frames $\{\mathcal{F}_1\}, \, \ldots ,\,\{\mathcal{F}_f\}$ related by $\boldsymbol{\mathcal{T}}_{\mathcal{A}\mathcal{F}_1}, \, \boldsymbol{\mathcal{T}}_{\mathcal{F}_1\mathcal{F}_2}, \, \ldots \,,\,\boldsymbol{\mathcal{T}}_{\mathcal{F}_f\mathcal{B}}$ to each other, and the frames $\{\mathcal{A}\}$ and $\{\mathcal{B}\}$, the relation
\begin{equation*}
    \boldsymbol{\mathcal{T}}_{\mathcal{A}\mathcal{B}} =
    \boldsymbol{\mathcal{T}}_{\mathcal{A}\mathcal{F}_1}
    \boldsymbol{\mathcal{T}}_{\mathcal{F}_1\mathcal{F}_2}
    \ldots
    \boldsymbol{\mathcal{T}}_{\mathcal{F}_f\mathcal{B}}
    \label{eq,TmatRec}
\end{equation*}
holds \cite{lynch2017modern}. With $\{\mathcal{S}\}$ denoting an inertial frame and $\{\mathcal{B}\}$ indicating a body-fixed frame, the linear velocity $\boldsymbol{v}_\mathcal{B}$ and the angular velocity $\boldsymbol{\omega}_\mathcal{B}$ of frame $\{\mathcal{B}\}$ measured from the inertial frame $\{\mathcal{S}\}$ but expressed in frame $\{\mathcal{B}\}$ can be obtained using the inverse and the first time derivative of the transformation matrix $\boldsymbol{\mathcal{T}}_{\mathcal{S}\mathcal{B}}$, \cite{lynch2017modern}
\begin{equation}
    \begin{bmatrix}
        [\boldsymbol{\omega}_\mathcal{B}] & \boldsymbol{v}_\mathcal{B} \\
        \boldsymbol{0} & 0
    \end{bmatrix}
    =
    \boldsymbol{\mathcal{T}}_{\mathcal{S}\mathcal{B}}^{-1}
    \boldsymbol{\dot{\mathcal{T}}}_{\mathcal{S}\mathcal{B}},
    \label{eq,bodytwist}
\end{equation}
considering that the notation $\left[.\right]$ denotes the skew-symmetric matrix representation of a vector. Assuming that the transformation matrix is written as a function of the augmented generalized coordinates $\boldsymbol{\hat{q}}(t)\in \mathbb{R}^{\hat{n}}$, which generally is a combination of the minimal and/or auxiliary coordinates, and using (\ref{eq,bodytwist}), the linear Jacobian $\boldsymbol{\mathcal{J}}_\mathcal{B}$ and angular Jacobian $\boldsymbol{\mathcal{L}}_\mathcal{B}$ of frame $\{\mathcal{B}\}$ are calculated as
\begin{equation}
\begin{split}
    \boldsymbol{v}_\mathcal{B} &= \boldsymbol{\mathcal{J}}_\mathcal{B}(\boldsymbol{\hat{q}})\,\boldsymbol{\dot{\hat{q}}} \\
    \boldsymbol{\omega}_\mathcal{B} &= \boldsymbol{\mathcal{L}}_\mathcal{B}(\boldsymbol{\hat{q}})\,\boldsymbol{\dot{\hat{q}}}.
    \label{eq,Jacob}
\end{split}
\end{equation}
Combining the linear and angular velocities in (\ref{eq,Jacob}) forms the twist vector
\begin{equation}
    \boldsymbol{\mathcal{V}}_\mathcal{B} =
    \begin{bmatrix}
        \boldsymbol{\omega}_\mathcal{B} \\
        \boldsymbol{v}_\mathcal{B}
    \end{bmatrix}
    \in se(3)
    \label{eq,twist}
\end{equation}
with $se(3)$ being the Lie algebra associated with $SE(3)$. Substituting (\ref{eq,Jacob}) into (\ref{eq,twist}) yields the spatial Jacobian
\begin{equation}
    \boldsymbol{\widehat{J}}_\mathcal{B} =
    \begin{bmatrix}
        \boldsymbol{\mathcal{L}}_\mathcal{B} \\
        \boldsymbol{\mathcal{J}}_\mathcal{B}
    \end{bmatrix}
    \label{eq,spatialJacobian}
\end{equation}
to be used as
\begin{equation}
    \boldsymbol{\mathcal{V}}_\mathcal{B} =
    \boldsymbol{\widehat{J}}_\mathcal{B}\boldsymbol{\dot{\hat{q}}}.
    \label{eq,twistJacobian}
\end{equation}
To find the twist of frame $\{\mathcal{A}\}$ from the known twist of frame $\{\mathcal{B}\}$, the relation \cite{lynch2017modern}
\begin{equation}
    \boldsymbol{\mathcal{V}}_\mathcal{A} =
    \boldsymbol{Ad}_{\boldsymbol{\mathcal{T}}_{\mathcal{A}\mathcal{B}}}\boldsymbol{\mathcal{V}}_\mathcal{B}
    \label{eq,twistrelation}
\end{equation}
is used. $\boldsymbol{Ad}_{\boldsymbol{\mathcal{T}}_{\mathcal{A}\mathcal{B}}}$ is the adjoint representation of the transform matrix $\boldsymbol{\mathcal{T}}_{\mathcal{A}\mathcal{B}}$, and it is defined as
\begin{equation*}
    \boldsymbol{Ad}_{\boldsymbol{\mathcal{T}}_{\mathcal{A}\mathcal{B}}} =
    \begin{bmatrix}
        \boldsymbol{R}_{\mathcal{A}\mathcal{B}} & \boldsymbol{0} \\
        \left[\boldsymbol{p}_{\mathcal{A}\mathcal{B}}\right] \boldsymbol{R}_{\mathcal{A}\mathcal{B}} & \boldsymbol{R}_{\mathcal{A}\mathcal{B}}
    \end{bmatrix}.
    \label{eq,adjoint}
\end{equation*}
From the twist relation (\ref{eq,twistrelation}) and the spatial Jacobian (\ref{eq,spatialJacobian}), the following is obtained:
\begin{equation}
    \boldsymbol{\widehat{J}}_\mathcal{A} =
    \boldsymbol{Ad}_{\boldsymbol{\mathcal{T}}_{\mathcal{A}\mathcal{B}}}\boldsymbol{\widehat{J}}_\mathcal{B}.
    \label{eq,Jacobianrelation}
\end{equation}

The spatial form of kinematics has provided a path to calculate the spatial Jacobian of frames from previously known spatial Jacobians. Also, it enables the use of the twist of any arbitrary body-fixed frame to be used in the calculation of the kinetic energy. In order to avoid the need for explicit constraint wrench calculations, we need to map the twists from the augmented generalized coordinate set $\boldsymbol{\hat{q}}$ to the minimal generalized coordinate set $\boldsymbol{q} \in \mathbb{R}^n$ where $n$ denotes the number of degrees of freedom. Assuming the mapping
\begin{equation*}
    \boldsymbol{\hat{q}} = \boldsymbol{\Xi}(\boldsymbol{q})
\end{equation*}
connects the minimal and augmented generalized coordinate sets, the intermediate Jacobian
\begin{equation*}
    \boldsymbol{E} = \dfrac{\partial \Xi}{\partial \boldsymbol{q}}
\end{equation*}
allows for mapping the twists to the minimal coordinate space. As a result, the spatial Jacobian $\boldsymbol{J}_\mathcal{A}$ in the minimal coordinates is obtained as
\begin{equation*}
    \boldsymbol{J}_\mathcal{A} = \boldsymbol{\widehat{J}}_\mathcal{A} \boldsymbol{E}.
\end{equation*}

\subsection{Generalized Inertia Matrix}
\label{subsec:inertiamatrix}

Kinetic energy of the $i^\text{th}$ rigid body in a multibody system can be calculated as
\begin{equation}
    T_i = \frac{1}{2}\boldsymbol{\mathcal{V}}_{\mathcal{B}_i}^T\boldsymbol{M}_i\boldsymbol{\mathcal{V}}_{\mathcal{B}_i}
    \label{eq,localkineticenergy}
\end{equation}
for an arbitrary body-fixed frame $\{\mathcal{B}_i\}$. The spatial inertia matrix $\boldsymbol{M}_i$ is defined as \cite{Featherstone2008}
\begin{equation*}
    \boldsymbol{M}_i =
    \begin{bmatrix}
        \boldsymbol{I}_{\mathcal{C}_i}+m_i\left[\boldsymbol{p}_{\mathcal{B}_i\mathcal{C}_i}\right]\left[\boldsymbol{p}_{\mathcal{B}_i\mathcal{C}_i}\right]^T & m_i\left[\boldsymbol{p}_{\mathcal{B}_i\mathcal{C}_i}\right] \\
        -m_i\left[\boldsymbol{p}_{\mathcal{B}_i\mathcal{C}_i}\right] & m_i \boldsymbol{\mathcal{I}}_3
    \end{bmatrix}
    \label{eq,spatialinertia}
\end{equation*}
with $m_i$ being the mass of the body and $\boldsymbol{I}_{\mathcal{C}_i}$ denoting the inertia tensor of the body with respect to frame $\{\mathcal{C}_i\}$ that coincides with the center of mass. The vector $\boldsymbol{p}_{\mathcal{B}_i\mathcal{C}_i}$ is the distance vector from the origin of frame $\{\mathcal{B}_i\}$ to the origin of frame $\{\mathcal{C}_i\}$ and expressed in $\{\mathcal{B}_i\}$. $\boldsymbol{\mathcal{I}}_3$ denotes the identity matrix of order 3.

By replacing the velocities in (\ref{eq,localkineticenergy}) with (\ref{eq,twistJacobian}), the kinetic energy of the body is rewritten as
\begin{equation*}
    T_i = \frac{1}{2}\boldsymbol{\dot{q}}^T \boldsymbol{\Gamma}_i(\boldsymbol{q})\boldsymbol{\dot{q}}.
    \label{eq,newlocalkinetic}
\end{equation*}
The matrix $\boldsymbol{\Gamma}_i$ is the generalized inertia matrix of the $i^\text{th}$ body and can be calculated as
\begin{equation}
    \boldsymbol{\Gamma}_i = \boldsymbol{J}_{\mathcal{B}_i}^T\boldsymbol{M}_i\boldsymbol{J}_{\mathcal{B}_i}.
    \label{eq,generalizedinertialmatrix}
\end{equation}

The local generalized inertia matrices do not represent a decomposition of the multibody system into detached physical subsystems. Instead, they quantify how the inertia of each body contributes to the motions associated with the degrees of freedom of the system, as represented by the chosen minimal generalized coordinates. In this sense, inertia partitioning refers to partitioning body-wise inertial effects with respect to coordinate-induced motion, as represented through the corresponding body-fixed-frame Jacobians. The global generalized inertia matrix is then obtained by assembling these body-level contributions in the common minimal-coordinate space. Since the total kinetic energy of the multibody system is the sum of the kinetic energies of the individual bodies, the global generalized inertia matrix $\boldsymbol{\Gamma}$ is obtained as the sum of the local generalized inertia matrices; thus,
\begin{equation}
    T = \sum_i T_i
    = \frac{1}{2}\boldsymbol{\dot{q}}^T \left(\sum_i\boldsymbol{\Gamma}_i(\boldsymbol{q})\right)\boldsymbol{\dot{q}}
    = \frac{1}{2}\boldsymbol{\dot{q}}^T \boldsymbol{\Gamma}(\boldsymbol{q})\boldsymbol{\dot{q}}.
    \label{eq,globalkinetic}
\end{equation}

Since the spatial inertia matrix of each rigid body is symmetric positive definite for physically meaningful mass and inertia parameters, each local generalized inertia matrix
\[
\boldsymbol{\Gamma}_i(\boldsymbol{q})=\boldsymbol{J}_{B_i}^T \boldsymbol{M}_i \boldsymbol{J}_{B_i}
\]
is symmetric positive semidefinite. In general, positive definiteness of an individual local generalized inertia matrix is not guaranteed, since the motion induced by a nonzero generalized velocity may not excite every body independently. However, for a physically admissible multibody system described by independent minimal generalized coordinates, the total kinetic energy is strictly positive for any nonzero generalized velocity. Therefore, the global generalized inertia matrix
\[
\boldsymbol{\Gamma}(\boldsymbol{q})=\sum_i \boldsymbol{\Gamma}_i(\boldsymbol{q})
\]
is symmetric positive definite on the admissible domain. Although the equations of motion are ultimately written in a single global form, their construction remains modular. The global generalized inertia matrix $\boldsymbol{\Gamma}(\boldsymbol{q})$ is assembled from body-level contributions, each obtained from the corresponding body Jacobian and spatial inertia matrix. As a result, modifications in the geometric or inertial properties of a body are reflected locally through the affected Jacobian or spatial inertia term, and the updated local contribution can then be reassembled into the global dynamics without re-deriving the entire model. Therefore, the generalized inertia matrix is obtained through modular body-level contributions while remaining expressed in the common minimal-coordinate space required for system-level dynamics and control. This provides the basis for deriving the equations of motion in the next part.

\subsection{Modular Equations of Motion}
With the global generalized inertia matrix assembled from the body-level contributions in \ref{subsec:inertiamatrix}, the equations of motion can now be derived in the minimal generalized coordinates using the Lagrangian formalism. For the present formulation, we choose to write the Lagrangian as the kinetic energy only, i.e., $(L=T)$, while the effects of gravity and energy-conserving elements such as springs are incorporated through the generalized force vector $\boldsymbol{Q}$. The generalized forces corresponding to the external wrenches $\boldsymbol{W}_k$ applied to the origin of frame $\{P_k\}$ can be calculated by \cite{Featherstone2008}
\begin{equation}
    \boldsymbol{Q} = \sum_k \boldsymbol{J}_{P_k}^{T}\boldsymbol{W}_k .
    \label{eq:genforces}
\end{equation}

Using the Lagrangian formalism \cite{shabana2009computational}
\begin{equation*}
    \frac{d}{dt}\left(\frac{\partial L}{\partial \dot{\boldsymbol{q}}}\right)-\frac{\partial L}{\partial \boldsymbol{q}}=\boldsymbol{Q},
    \label{eq:lagrangeformalism}
\end{equation*}
and using \eqref{eq,globalkinetic} in accordance with the mentioned assumptions, the equations of motion for the whole system are achieved as
\begin{equation}
    \boldsymbol{\Gamma}(\boldsymbol{q})\ddot{\boldsymbol{q}}
    + \dot{\boldsymbol{\Gamma}}(\boldsymbol{q},\dot{\boldsymbol{q}})\dot{\boldsymbol{q}}
    - \frac{1}{2}\nabla_{\boldsymbol{q}}
    \left(\dot{\boldsymbol{q}}^{T}\boldsymbol{\Gamma}(\boldsymbol{q})\dot{\boldsymbol{q}}\right)
    = \boldsymbol{Q}.
    \label{eq:globalEOM}
\end{equation}

The first time derivative of the generalized inertia matrix can be obtained using the chain rule
\begin{equation*}
    \dot{\boldsymbol{\Gamma}}
    =
    \sum_j
    \left(
        \frac{\partial \boldsymbol{\Gamma}}{\partial q_j}\dot{q}_j
    \right).
    \label{eq:gammadotchain}
\end{equation*}

The modular equations of motion obtained in this part will be used in the controller design in Section \ref{sec:control}.


\section{Modular Control of Multibody Systems}
\label{sec:control}

In this section, the modular equations of motion are used to design a stable model-based controller for the multibody system. Although the final control law is written in a global form, its model-based component is constructed from the globally assembled dynamics whose terms originate from local body-level contributions. Therefore, the benefit of modularity in the present framework lies in systematic model construction and updatability, rather than in decentralized controller synthesis.

\subsection{Controller Design}

The generalized force $\boldsymbol{Q}$ is separated into two parts, namely the generalized external force $\boldsymbol{Q}_e$ and the generalized force $\boldsymbol{Q}_a$ generated by the actuators, such that
\begin{equation*}
    \boldsymbol{Q} = \boldsymbol{Q}_e + \boldsymbol{Q}_a .
    \label{eq:Qsplit}
\end{equation*}
Using the modular equation of motion in \eqref{eq:globalEOM}, the multibody system dynamics can be written as
\begin{equation}
    \boldsymbol{\Gamma}(\boldsymbol{q})\ddot{\boldsymbol{q}}
    + \boldsymbol{h}(\boldsymbol{q},\dot{\boldsymbol{q}})
    =
    \boldsymbol{Q}_e(\boldsymbol{q},\dot{\boldsymbol{q}})
    + \boldsymbol{Q}_a ,
    \label{eq:plantcontrol}
\end{equation}
where
\begin{equation*}
    \boldsymbol{h}(\boldsymbol{q},\dot{\boldsymbol{q}})
    =
    \dot{\boldsymbol{\Gamma}}(\boldsymbol{q},\dot{\boldsymbol{q}})
    \dot{\boldsymbol{q}}
    - \frac{1}{2}
    \nabla_{\boldsymbol{q}}
    \left(
        \dot{\boldsymbol{q}}^T
        \boldsymbol{\Gamma}(\boldsymbol{q})
        \dot{\boldsymbol{q}}
    \right).
    \label{eq:htermcontrol}
\end{equation*}

Let the desired trajectory and its derivatives be denoted by $\boldsymbol{q}_d$, $\dot{\boldsymbol{q}}_d$, and $\ddot{\boldsymbol{q}}_d$, respectively. Define the tracking error as
\begin{equation}
    \boldsymbol{e} = \boldsymbol{q} - \boldsymbol{q}_d,
    \qquad
    \dot{\boldsymbol{e}} = \dot{\boldsymbol{q}} - \dot{\boldsymbol{q}}_d .
    \label{eq:trackingerror}
\end{equation}
To shape the tracking error dynamics, we define the filtered tracking variable
\begin{equation}
    \boldsymbol{s} = \dot{\boldsymbol{e}} + \boldsymbol{\Lambda}\boldsymbol{e},
    \label{eq:filterederror}
\end{equation}
where $\boldsymbol{\Lambda}$ is a diagonal positive definite matrix. Also, define the reference velocity and acceleration as
\begin{equation}
    \dot{\boldsymbol{q}}_r
    =
    \dot{\boldsymbol{q}}_d
    -
    \boldsymbol{\Lambda}\boldsymbol{e},
    \qquad
    \ddot{\boldsymbol{q}}_r
    =
    \ddot{\boldsymbol{q}}_d
    -
    \boldsymbol{\Lambda}\dot{\boldsymbol{e}}.
    \label{eq:refsignals}
\end{equation}
Using \eqref{eq:trackingerror} and \eqref{eq:refsignals}, it follows that
\begin{equation}
    \boldsymbol{s}
    =
    \dot{\boldsymbol{q}}-\dot{\boldsymbol{q}}_r ,
    \qquad
    \dot{\boldsymbol{s}}
    =
    \ddot{\boldsymbol{q}}-\ddot{\boldsymbol{q}}_r .
    \label{eq:srelations}
\end{equation}

The model-based part of the control law is obtained from \eqref{eq:plantcontrol} by replacing $\ddot{\boldsymbol{q}}$ with the reference acceleration $\ddot{\boldsymbol{q}}_r$. Then, linear error feedback and robust correction terms are added on top of the model-based part to propose the control law
\begin{equation}
    \boldsymbol{Q}_a
    =
    \boldsymbol{\Gamma}(\boldsymbol{q})\ddot{\boldsymbol{q}}_r
    +
    \boldsymbol{h}(\boldsymbol{q},\dot{\boldsymbol{q}})
    -
    \boldsymbol{Q}_e(\boldsymbol{q},\dot{\boldsymbol{q}})
    -
    \boldsymbol{K}\boldsymbol{s}
    -
    \boldsymbol{\kappa}
    \,\mathrm{sat}\!\left(
        \frac{\boldsymbol{s}}{\boldsymbol{\varepsilon}}
    \right),
    \label{eq:controller}
\end{equation}
where $\boldsymbol{K}$ and $\boldsymbol{\kappa}$ are diagonal positive definite matrices. $\boldsymbol{\varepsilon}$ is a vector of small positive scalars, and the saturation function and the division in its argument are applied component-wise.

The model-based part in \eqref{eq:controller} is directly obtained from the modularly assembled equations of motion, while the correction terms are used to shape the error dynamics and to improve robustness against modeling uncertainties and external disturbances.

\subsection{Practical Stability Analysis Under Bounded Uncertainty}

To analyze the robustness of the proposed controller in the presence of bounded uncertainty, consider the actual system
\begin{equation}
    \boldsymbol{\Gamma}(\boldsymbol{q})\ddot{\boldsymbol{q}}
    +
    \boldsymbol{h}(\boldsymbol{q},\dot{\boldsymbol{q}})
    =
    \boldsymbol{Q}_e(\boldsymbol{q},\dot{\boldsymbol{q}})
    +
    \boldsymbol{Q}_a
    +
    \boldsymbol{d}(\boldsymbol{q},\dot{\boldsymbol{q}},t),
    \label{eq:uncertainplant}
\end{equation}
where $\boldsymbol{d} \in \mathbb{R}^n$ denotes a lumped uncertainty term accounting for modeling errors, parameter mismatch, and unmodeled effects.

Substituting \eqref{eq:controller} into \eqref{eq:uncertainplant} and using \eqref{eq:srelations} gives
\begin{equation}
    \boldsymbol{\Gamma}(\boldsymbol{q})\dot{\boldsymbol{s}}
    +
    \boldsymbol{K}\boldsymbol{s}
    +
    \boldsymbol{\kappa}
    \,\mathrm{sat}\!\left(
        \frac{\boldsymbol{s}}{\boldsymbol{\varepsilon}}
    \right)
    =
    \boldsymbol{d}.
    \label{eq:sdynbasic}
\end{equation}

For the Lyapunov analysis, \eqref{eq:uncertainplant} is written in the standard manipulator form
\begin{equation*}
    \boldsymbol{\Gamma}(\boldsymbol{q})\ddot{\boldsymbol{q}}
    +
    \boldsymbol{C}(\boldsymbol{q},\dot{\boldsymbol{q}})
    \dot{\boldsymbol{q}}
    =
    \boldsymbol{Q}_e(\boldsymbol{q},\dot{\boldsymbol{q}})
    +
    \boldsymbol{Q}_a
    +
    \boldsymbol{d},
    \label{eq:manipulatorform}
\end{equation*}
where $\boldsymbol{C}(\boldsymbol{q},\dot{\boldsymbol{q}})$ can be chosen such that
\begin{equation*}
    \dot{\boldsymbol{\Gamma}}(\boldsymbol{q},\dot{\boldsymbol{q}})
    -
    2\boldsymbol{C}(\boldsymbol{q},\dot{\boldsymbol{q}})
\end{equation*}
is skew-symmetric \cite{murray1994mathematical}. Using \eqref{eq:srelations}, the filtered error dynamics become
\begin{equation}
    \boldsymbol{\Gamma}(\boldsymbol{q})\dot{\boldsymbol{s}}
    +
    \boldsymbol{C}(\boldsymbol{q},\dot{\boldsymbol{q}})
    \boldsymbol{s}
    +
    \boldsymbol{K}\boldsymbol{s}
    +
    \boldsymbol{\kappa}
    \,\mathrm{sat}\!\left(
        \frac{\boldsymbol{s}}{\boldsymbol{\varepsilon}}
    \right)
    =
    \boldsymbol{d}.
    \label{eq:sdynCform}
\end{equation}

Consider the Lyapunov candidate
\begin{equation}
    V
    =
    \frac{1}{2}
    \boldsymbol{s}^T
    \boldsymbol{\Gamma}(\boldsymbol{q})
    \boldsymbol{s}.
    \label{eq:lyapnew}
\end{equation}
Since $\boldsymbol{\Gamma}(\boldsymbol{q})$ is symmetric positive definite, there exist positive constants $\underline{\gamma}$ and $\overline{\gamma}$ such that
\begin{equation}
    \frac{1}{2}\underline{\gamma}\|\boldsymbol{s}\|^2
    \le
    V
    \le
    \frac{1}{2}\overline{\gamma}\|\boldsymbol{s}\|^2
    \label{eq:Vbounds}
\end{equation}
on any compact admissible domain.

Differentiating \eqref{eq:lyapnew} and substituting \eqref{eq:sdynCform} yields
\begin{equation*}
    \dot{V}
    =
    -
    \boldsymbol{s}^T
    \boldsymbol{K}
    \boldsymbol{s}
    -
    \boldsymbol{s}^T
    \boldsymbol{\kappa}
    \,\mathrm{sat}\!\left(
        \frac{\boldsymbol{s}}{\boldsymbol{\varepsilon}}
    \right)
    +
    \boldsymbol{s}^T\boldsymbol{d}
    +
    \frac{1}{2}
    \boldsymbol{s}^T
    \left(
        \dot{\boldsymbol{\Gamma}}
        -
        2\boldsymbol{C}
    \right)
    \boldsymbol{s}.
    \label{eq:Vdotstep2}
\end{equation*}
The last term vanishes by skew-symmetry, and therefore
\begin{equation}
    \dot{V}
    =
    -
    \boldsymbol{s}^T
    \boldsymbol{K}
    \boldsymbol{s}
    -
    \boldsymbol{s}^T
    \boldsymbol{\kappa}
    \,\mathrm{sat}\!\left(
        \frac{\boldsymbol{s}}{\boldsymbol{\varepsilon}}
    \right)
    +
    \boldsymbol{s}^T\boldsymbol{d}.
    \label{eq:Vdotfinal}
\end{equation}

Assume that the uncertainty is bounded componentwise as
\begin{equation*}
    |d_i(\boldsymbol{q},\dot{\boldsymbol{q}},t)|
    \le
    \bar{d}_i,
    \qquad
    i=1,\ldots,n,
    \label{eq:dbound}
\end{equation*}
and choose the robust gains such that
\begin{equation}
    \kappa_i > \bar{d}_i,
    \qquad
    i=1,\ldots,n.
    \label{eq:kappacond}
\end{equation}
Using
\begin{equation*}
    s_i\,\mathrm{sat}\!\left(\frac{s_i}{\varepsilon_i}\right)
    \ge
    |s_i|-\varepsilon_i,
    \label{eq:satineq}
\end{equation*}
it follows from \eqref{eq:Vdotfinal} that
\begin{equation*}
    \dot{V}
    \le
    -
    \lambda_{\min}(\boldsymbol{K})
    \|\boldsymbol{s}\|^2
    -
    \sum_{i=1}^{n}
    (\kappa_i-\bar{d}_i)|s_i|
    +
    \sum_{i=1}^{n}\kappa_i \varepsilon_i .
    \label{eq:Vdotineq}
\end{equation*}

Hence, $\dot{V}$ is negative outside a compact neighborhood of the origin whose size depends on the uncertainty bounds $\bar d_i$ and the boundary-layer widths $\varepsilon_i$. Together with \eqref{eq:Vbounds}, this implies that the filtered tracking error $\boldsymbol{s}$ is uniformly ultimately bounded \cite{Khalil2002NonlinearSystems}. Since \eqref{eq:filterederror} is a stable linear filter driven by $\boldsymbol{s}$, the tracking error $\boldsymbol{e}$ is also uniformly ultimately bounded. Therefore, under bounded uncertainty and \eqref{eq:kappacond}, the proposed controller guarantees practical boundedness of both $\boldsymbol{s}$ and $\boldsymbol{e}$.


\section{Simulation and Results}
\label{sec:simulation}

To assess the performance of the robust inertia partitioning framework for multibody systems with closed kinematic chains, a simulation is conducted on a 3-DoF series-parallel manipulator using Simscape, the physics engine of MATLAB.

\begin{figure}[t]
    \centering
    \begin{minipage}[h]{0.45\textwidth}
        \includegraphics[width=\linewidth,trim={5cm, 1cm, 2cm, 0cm},clip]{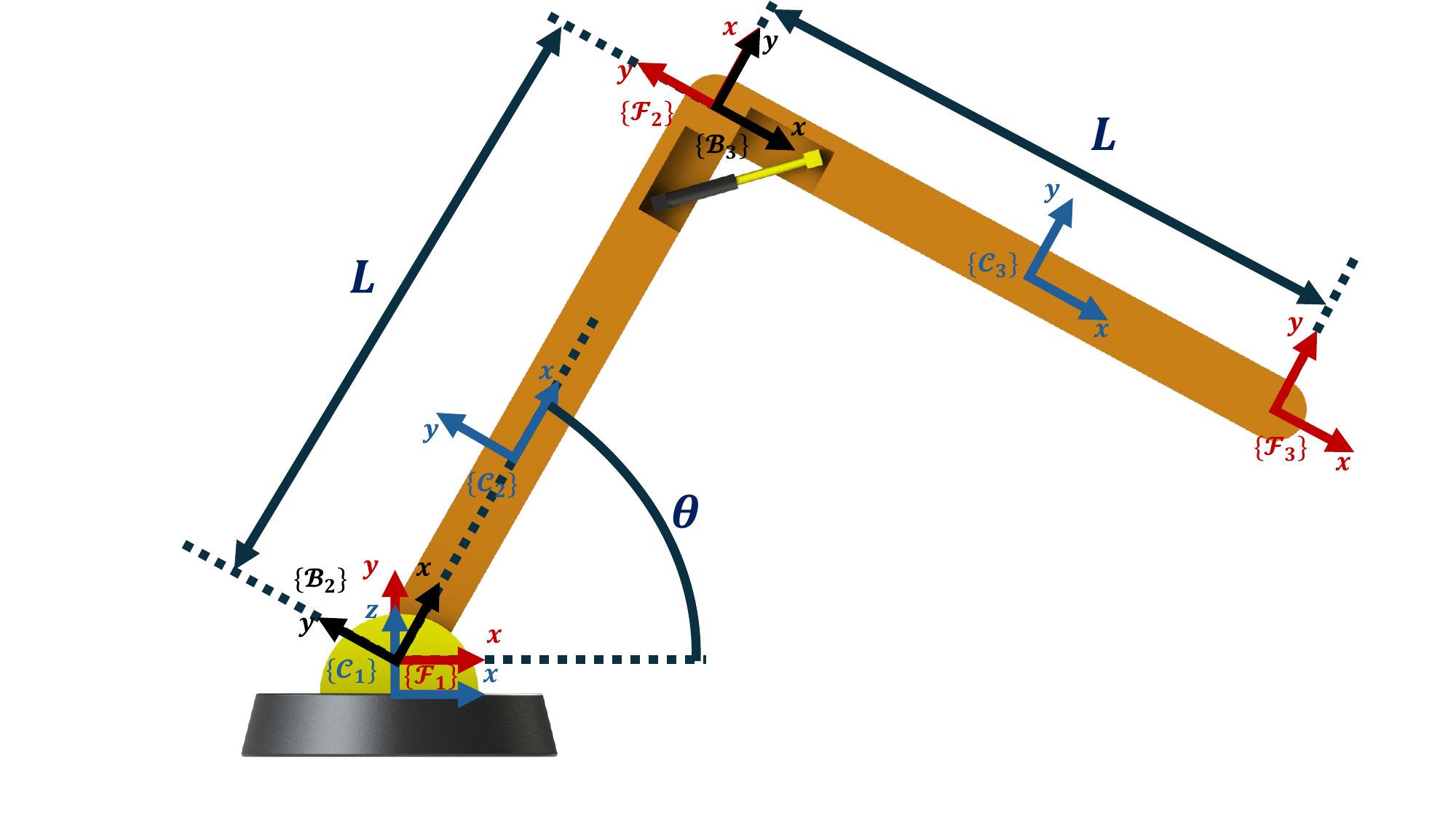}
        \label{side}
    \end{minipage}
    \hfill
    \begin{minipage}[h]{0.45\textwidth}
        \includegraphics[width=\linewidth,trim={7cm, 4cm, 7cm, 3cm},clip]{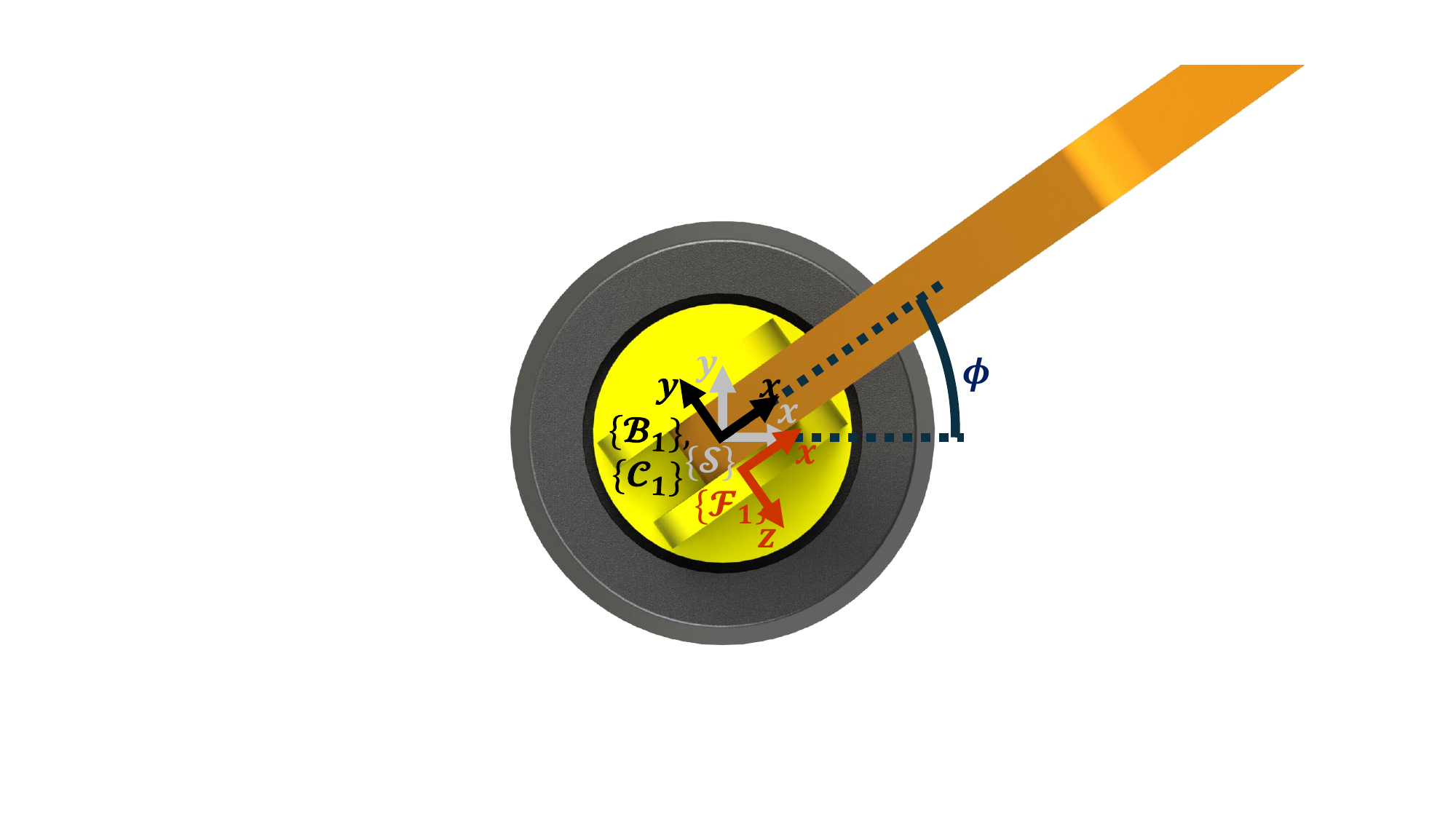}
        \label{top}
    \end{minipage}
    \hfill
    \caption{The schematics for a 3-DoF series-parallel manipulator. The degrees of freedom are made up of two actuated revolute joints for the base and the first link, and an actuated prismatic joint between the first and second links. The required frames of reference and geometric parameters for modeling the 3-DoF series-parallel manipulator. Some frames are shown only in one of the figures for clarity. \textit{Top} - Side view perpendicular to the links' body-fixed frames $x$-$y$ plane. \textit{Bottom} - Top view of the base, perpendicular to its frames' $x$-$y$ planes.}
    \label{frames}
\end{figure}

The series-parallel manipulator in the simulation is shown in Fig. \ref{frames}. The frames of reference needed for modeling this multibody system, as well as the required geometric parameters, are illustrated in the same figure. The orientation of frames is chosen such that they comply with the Simscape convention for mechanical joints. Frame $\{\mathcal{S}\}$ is the inertial frame of reference, while the other frames are body-fixed. Indices $1, 2,$ and $3$ refer to body-fixed frames attached to the base, the first link, and the second link, respectively.

\begin{figure}[t]
    \centering
    \includegraphics[width=0.75\linewidth,trim={5cm, 1cm, 5cm, 1.5cm},clip]{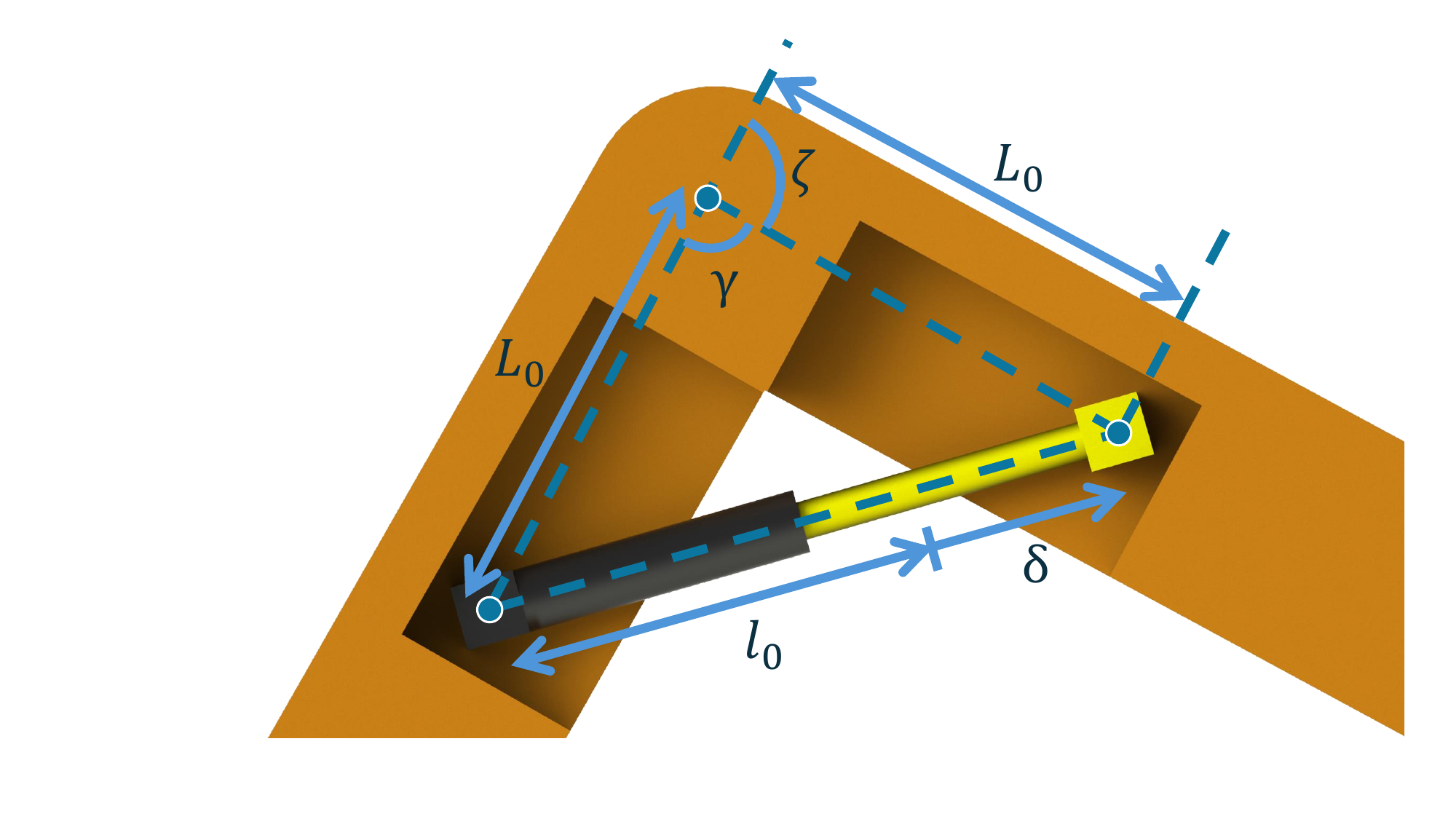}
    \caption{The geometry of the triangular closed kinematic chain. This closed kinematic chain consists of one actuated prismatic joint and three passive revolute joints. In this closed kinematic chain, the prismatic joint has the limit of $0.2\;m$ for $\delta$}
    \label{triangle}
\end{figure}

The generalized coordinates used to model this series-parallel manipulator are
\begin{equation*}
    \boldsymbol{q}=
    \begin{bmatrix}
        \phi & \theta & \delta
    \end{bmatrix}^T,
    \label{eq:qsim}
\end{equation*}
as depicted in Figs. \ref{frames} and \ref{triangle}. To express the relative rotation between the first and the second links, the intermediate coordinate $\zeta$, which is a function of $\delta$, is also introduced. Accordingly, for the purpose of expressing the kinematics through the intermediate coordinate $\zeta$, we introduce the coordinate set
\[
\boldsymbol{\hat{q}} = \begin{bmatrix}\phi & \theta & \zeta\end{bmatrix}^T,
\]
which is related to the minimal generalized coordinates
\[
\boldsymbol{q} = \begin{bmatrix}\phi & \theta & \delta\end{bmatrix}^T
\]
through the mapping $\zeta=\zeta(\delta)$. The corresponding intermediate Jacobian is
\begin{equation}
    \boldsymbol{E}=
    \frac{\partial \boldsymbol{\hat{q}}}{\partial \boldsymbol{q}}
    =
    \begin{bmatrix}
        1 & 0 & 0 \\
        0 & 1 & 0 \\
        0 & 0 & \dfrac{\partial \zeta}{\partial \delta}
    \end{bmatrix}.
    \label{eq:Eexample}
\end{equation}

Using the twist relation (\ref{eq,twistJacobian}) and the Jacobian transformation (\ref{eq,Jacobianrelation}), while considering the mechanical joints in the 3-DoF series-parallel manipulator, the Jacobians of the required frames are calculated as
\begin{equation}
\begin{split}
    \boldsymbol{J}_{\mathcal{B}_1}
    &=
    \begin{bmatrix}
        0 & 0 & 1 & 0 & 0 & 0 \\
        0 & 0 & 0 & 0 & 0 & 0 \\
        0 & 0 & 0 & 0 & 0 & 0
    \end{bmatrix}^T \\
    \boldsymbol{J}_{\mathcal{C}_1}
    &=
    \boldsymbol{Ad}_{\boldsymbol{\mathcal{T}}_{\mathcal{C}_1\mathcal{B}_1}}\boldsymbol{J}_{\mathcal{B}_1} \\
    \boldsymbol{J}_{\mathcal{F}_1}
    &=
    \boldsymbol{Ad}_{\boldsymbol{\mathcal{T}}_{\mathcal{F}_1\mathcal{B}_1}}\boldsymbol{J}_{\mathcal{B}_1} \\
    \boldsymbol{J}_{\mathcal{B}_2}
    &=
    \boldsymbol{Ad}_{\boldsymbol{\mathcal{T}}_{\mathcal{B}_2\mathcal{F}_1}}\boldsymbol{J}_{\mathcal{F}_1}
    +
    \begin{bmatrix}
        0 & 0 & 0 & 0 & 0 & 0 \\
        0 & 0 & 1 & 0 & 0 & 0 \\
        0 & 0 & 0 & 0 & 0 & 0
    \end{bmatrix}^T \\
    \boldsymbol{J}_{\mathcal{C}_2}
    &=
    \boldsymbol{Ad}_{\boldsymbol{\mathcal{T}}_{\mathcal{C}_2\mathcal{B}_2}}\boldsymbol{J}_{\mathcal{B}_2} \\
    \boldsymbol{J}_{\mathcal{F}_2}
    &=
    \boldsymbol{Ad}_{\boldsymbol{\mathcal{T}}_{\mathcal{F}_2\mathcal{B}_2}}\boldsymbol{J}_{\mathcal{B}_2} \\
    \boldsymbol{\widehat{J}}_{\mathcal{B}_3}
    &=
    \boldsymbol{Ad}_{\boldsymbol{\mathcal{T}}_{\mathcal{B}_3\mathcal{F}_2}}\boldsymbol{J}_{\mathcal{F}_2}
    +
    \begin{bmatrix}
        0 & 0 & 0 & 0 & 0 & 0 \\
        0 & 0 & 0 & 0 & 0 & 0 \\
        0 & 0 & 1 & 0 & 0 & 0
    \end{bmatrix}^T \\
    \boldsymbol{J}_{\mathcal{B}_3}
    &=
    \boldsymbol{\widehat{J}}_{\mathcal{B}_3}\boldsymbol{E} \\
    \boldsymbol{J}_{\mathcal{C}_3}
    &=
    \boldsymbol{Ad}_{\boldsymbol{\mathcal{T}}_{\mathcal{C}_3\mathcal{B}_3}}\boldsymbol{J}_{\mathcal{B}_3} \\
    \boldsymbol{J}_{\mathcal{F}_3}
    &=
    \boldsymbol{Ad}_{\boldsymbol{\mathcal{T}}_{\mathcal{F}_3\mathcal{B}_3}}\boldsymbol{J}_{\mathcal{B}_3}.
\end{split}
    \label{eq,simJacobian}
\end{equation}

The rotation matrices between the frames are obtainable from Figs. \ref{frames} and \ref{triangle}, and the non-zero required distance vectors are available in Table \ref{tab,dist}. In order to calculate the intermediate Jacobian $\boldsymbol{E}$, a geometric relation is needed between the intermediate coordinate $\zeta$ and the degree-of-freedom coordinate $\delta$. The triangular closed kinematic chain between the first and second links is shown more closely in Fig. \ref{triangle}. Applying the law of cosines for angle $\gamma$ yields
\begin{equation*}
    \cos(\gamma)=1-\dfrac{(l_0+\delta)^2}{2L_0^2}
    \label{eq,gamma}
\end{equation*}
with $l_0$ being the length of the cylinder-piston pair when the piston is at its lower end, i.e., $\delta=0$. Because $\gamma$ and $\zeta$ are supplementary angles, and considering the fact that the value of rotation $\zeta$ with respect to the $z$-axis of frame $\{\mathcal{B}_3\}$ lies in a subset of the interval $(-\pi, 0)$ whereas the inverse cosine range lies in the interval $(0,\pi)$, the relationship between $\zeta$ and $\delta$ is obtained as
\begin{equation*}
    \zeta=-\cos^{-1}\left(\dfrac{(\delta+l_0)^2}{2L_0^2}-1\right),
    \label{eq,zeta}
\end{equation*}
and as a result
\begin{equation}
    \frac{\partial\zeta}{\partial\delta}=
    \dfrac{\delta+l_0}{L_0^2\sqrt{1-\left(\dfrac{(\delta+l_0)^2}{2L_0^2}-1\right)^2}}\;.
    \label{eq,zetadelta}
\end{equation}

\begin{table}[b]
    \centering
    \caption{Required non-zero distance vectors between the frames in the 3-DoF series-parallel manipulator. All values are in meters.}
    \begin{tabular}{|r@{\,=\,}l|}
        \hline
        \rule{0pt}{10pt}$\boldsymbol{p}_{\mathcal{C}_1\mathcal{B}_1}$ & $\begin{bmatrix} 0 & 0 & -0.103 \end{bmatrix}^T$ \\[2pt]
        \hline
        \rule{0pt}{10pt}$\boldsymbol{p}_{\mathcal{F}_1\mathcal{B}_1}$ & $\begin{bmatrix} 0 & -0.206 & -0.075 \end{bmatrix}^T$ \\[2pt]
        \hline
        \rule{0pt}{10pt}$\boldsymbol{p}_{\mathcal{C}_2\mathcal{B}_2}$ & $\begin{bmatrix} -0.959 & -0.001 & 0.077 \end{bmatrix}^T$ \\[2pt]
        \hline
        \rule{0pt}{10pt}$\boldsymbol{p}_{\mathcal{F}_2\mathcal{B}_2}$ & $\begin{bmatrix} -2 & 0 & 0 \end{bmatrix}^T$ \\[2pt]
        \hline
        \rule{0pt}{10pt}$\boldsymbol{p}_{\mathcal{C}_3\mathcal{B}_3}$ & $\begin{bmatrix} -1.041 & -0.001 & 0.077 \end{bmatrix}^T$ \\[2pt]
        \hline
        \rule{0pt}{10pt}$\boldsymbol{p}_{\mathcal{F}_3\mathcal{B}_3}$ & $\begin{bmatrix} -2 & 0 & 0 \end{bmatrix}^T$ \\[2pt]
        \hline
    \end{tabular}
    \label{tab,dist}
\end{table}

With the Jacobians obtained from (\ref{eq:Eexample}), (\ref{eq,simJacobian}), and (\ref{eq,zetadelta}), the generalized inertia matrix for each body is obtained using (\ref{eq,generalizedinertialmatrix}), with their summation yielding the global generalized inertia matrix $\boldsymbol{\Gamma}$. To introduce unmodeled dynamics in the simulation, the inertial effects of the cylinder-piston pair are neglected in deriving the model-based part of the controller. The parameter values in the trajectory tracking simulation of the 3-DoF series-parallel manipulator are expressed in Table \ref{tab,param}. 

\begin{table}[t]
\caption{Physical and control parameter values for the 3-DoF series-parallel manipulator}
\centering
\renewcommand{\arraystretch}{1.0}
\setlength{\tabcolsep}{4pt}
\begin{tabular}{|l|c|}
\hline
Parameter & Value \\
\hline
$L$ & $2\,\mathrm{m}$ \\
\hline
$L_0$ & $0.35\,\mathrm{m}$ \\
\hline
$l_0$ & $0.425\,\mathrm{m}$ \\
\hline
$m_1$ & $30\,\mathrm{kg}$ \\
\hline
$m_2$ & $60\,\mathrm{kg}$ \\
\hline
$m_3$ & $60\,\mathrm{kg}$ \\
\hline
cylinder & $1.5\,\mathrm{kg}$ \\
\hline
piston & $1\,\mathrm{kg}$ \\
\hline
$\boldsymbol{I}_{\mathcal{C}_1}$ &
$\mathrm{diag}([0.804\;\;0.831\;\;1.18])\,\mathrm{kg\,m^2}$ \\
\hline
$\boldsymbol{I}_{\mathcal{C}_2}$ &
$\begin{bmatrix}
0.311 & -0.065 & 0.098 \\
-0.065 & 22.7 & -0.003 \\
0.098 & -0.003 & 22.8
\end{bmatrix}\,\mathrm{kg\,m^2}$ \\
\hline
$\boldsymbol{I}_{\mathcal{C}_3}$ &
$\begin{bmatrix}
0.311 & 0.065 & -0.098 \\
0.065 & 22.7 & -0.003 \\
-0.098 & -0.003 & 22.8
\end{bmatrix}\,\mathrm{kg\,m^2}$ \\
\hline
$g$ & $9.81\,\mathrm{m/s^2}$ \\
\hline
$\boldsymbol{\varepsilon}$ & $[10^{-4}\;\;10^{-4}\;\;10^{-4}]^T$ \\
\hline
$\boldsymbol{\Lambda}$ & $\mathrm{diag}([1\;\;1\;\;1])$ \\
\hline
$\boldsymbol{K}$ & $\mathrm{diag}([30\;\;30\;\;30])$ \\
\hline
$\boldsymbol{\kappa}$ & $\mathrm{diag}([50\;\;50\;\;50])$ \\
\hline
\end{tabular}
\label{tab,param}
\end{table}

To calculate the generalized forces corresponding to the weight of the bodies, the Jacobians $\boldsymbol{J}_{\mathcal{C}_i}$ and the rotation matrices $\boldsymbol{R}_{\mathcal{C}_i\mathcal{S}}$ of body-fixed frames $\{\mathcal{C}_i\}$ coincident with the centers of mass are needed. With those matrices available and considering the weight to be an external wrench, under the assumption of no friction in the joints, the corresponding generalized forces can be calculated by using (\ref{eq:genforces})
\begin{equation*}
    \boldsymbol{Q}=\sum_i \boldsymbol{J}_{\mathcal{C}_i}^T \boldsymbol{W}_i
    =
    \sum_i\boldsymbol{J}_{\mathcal{C}_i}^T
    \begin{bmatrix}
        \boldsymbol{R}_{\mathcal{C}_i\mathcal{S}} & \boldsymbol{0} \\
        \boldsymbol{0} & \boldsymbol{R}_{\mathcal{C}_i\mathcal{S}}
    \end{bmatrix}
    \begin{bmatrix}
        \boldsymbol{0} \\
        -m_ig
    \end{bmatrix}.
    \label{eq,weight}
\end{equation*}
With all the required components of the control law (\ref{eq:controller}) available either by calculation or by selection, the chosen 3-DoF series-parallel manipulator can be controlled for trajectory tracking.

\begin{figure}[t]
    \centering
    \includegraphics[width=\linewidth]{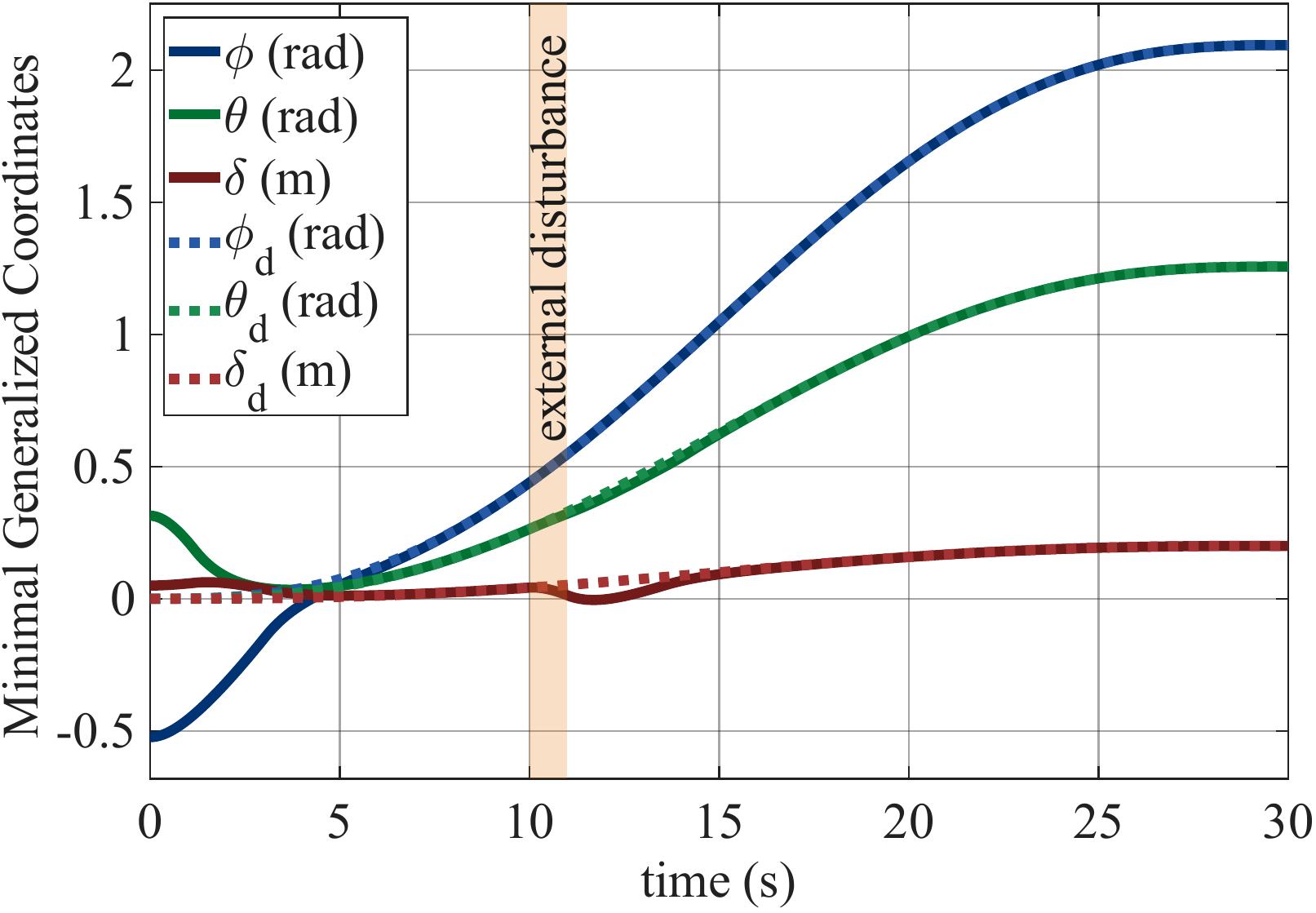}
    \caption{Trajectory tracking results for the 3-DoF series-parallel manipulator. Comparison of desired and actual trajectories with respect to time.}
    \label{track}
\end{figure}

\begin{figure}[t]
    \centering
    \includegraphics[width=\linewidth]{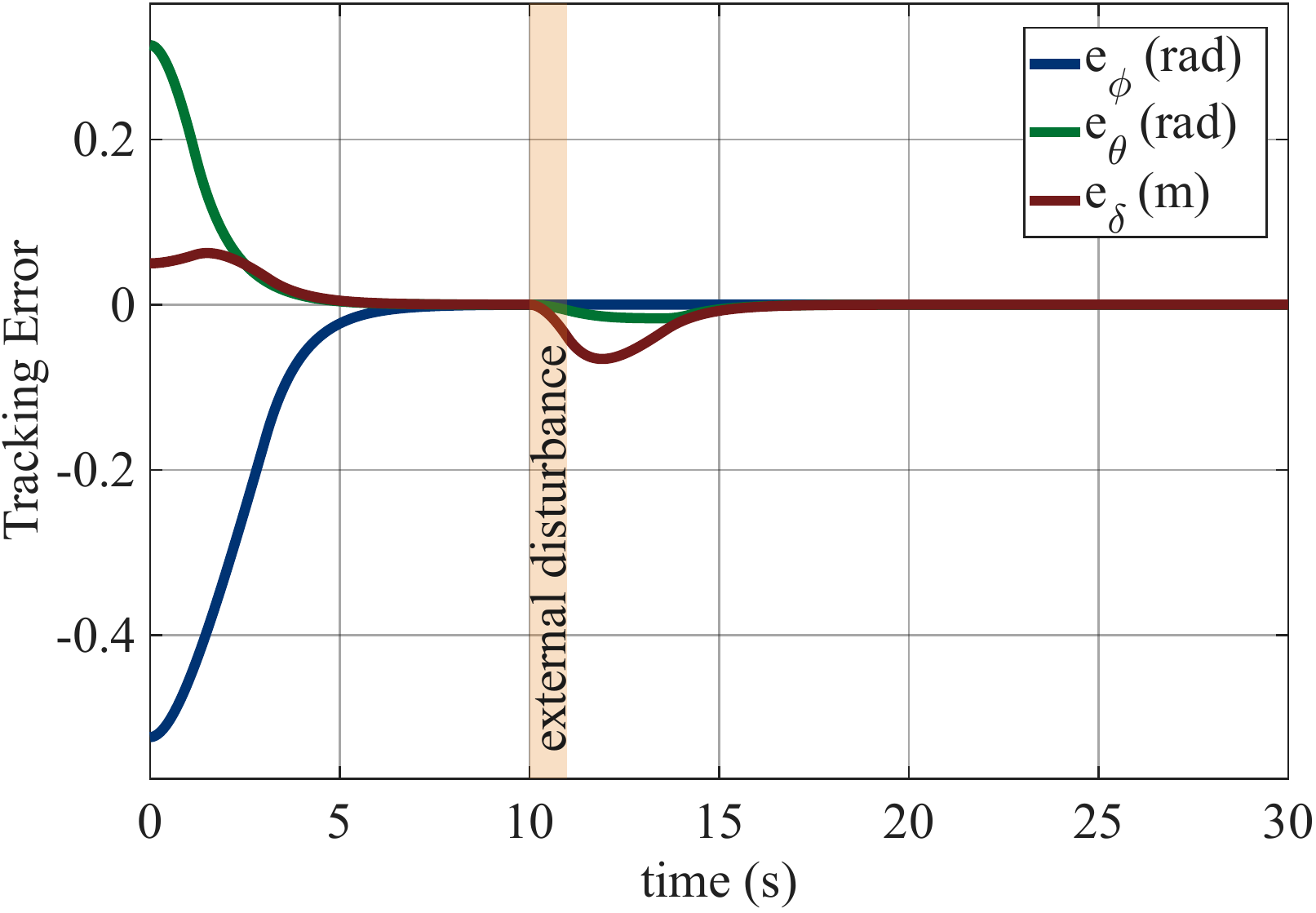}
    \caption{Trajectory tracking error. The controller was able to maintain the desired trajectory despite the external disturbance and unmodeled dynamics on top of the initial error values, supporting the robustness of the proposed controller under the tested conditions.}
    \label{error}
\end{figure}

The desired trajectory is defined in the minimal generalized coordinates $\boldsymbol{q}_d=[\phi_d \quad \theta_d \quad \delta_d]^T$, together with its first and second time derivatives required by the control law. In general, such reference trajectories may also be obtained by mapping a desired end-effector trajectory from Cartesian space into the minimal-coordinate representation. With the desired trajectory determined and the parameter values available in Table \ref{tab,param}, the control law (\ref{eq:controller}) is applied to the manipulator with initial error values in the minimal generalized coordinates. Also, a constant force with the magnitude of $25\,N$ is applied as an external disturbance in the negative $y-$direction of the frame $\{\mathcal{F}_3\}$ between the 10$^\text{th}$ and the 11$^\text{th}$ seconds. This is in addition to the previously mentioned unmodeled inertial effects of the piston-cylinder pair. Figs. \ref{track} and \ref{error} illustrate the trajectory tracking results. Fig. \ref{track} shows the followed trajectory against the desired one in the minimal generalized coordinates. Fig. \ref{error} displays the tracking error with respect to time using the error values in the minimal generalized coordinates. The root mean squared error (RMSE) values for the tracking error in the minimal generalized coordinates are reported in Table \ref{tab,rmse}.

\begin{table}[b]
    \caption{Root mean squared error values for trajectory tracking}
    \centering
    \begin{tabular}{|c|c|c|c|}
        \hline
        \rule{0pt}{10pt}  & $\phi$ & $\theta$ & $\delta$  \\[2pt]
        \hline
        \rule{0pt}{10pt}RMSE & 0.124 rad & 0.074 rad & 0.018 m \\[2pt]
        \hline
    \end{tabular}
    \label{tab,rmse}
\end{table}


\section{Conclusion}
\label{sec:conclusion}

The challenges associated with handling closed kinematic chains in modular control of multibody systems with existing methods motivated the development of the proposed modular control framework. Provided that the mapping between the auxiliary and minimal generalized coordinates is available, the framework accommodates closed kinematic chains in a minimal-coordinate form. The proposed framework defines modularity with respect to the degrees of freedom of the multibody system, represented in the model by the minimal generalized coordinates, and constructs the global dynamics through assembly of body-level inertial contributions expressed by body-fixed-frame Jacobians and spatial inertia matrices. In this sense, the method does not decompose the system by detaching physical bodies into independently controlled subsystems, but rather partitions how each body's inertia contributes to coordinate-induced motion in the common minimal-coordinate space. This provides a systematic and control-oriented route to obtaining the equations of motion for multibody systems, including systems with closed kinematic chains when the mapping between auxiliary and minimal generalized coordinates is available.

An important feature of the proposed framework is that reconfiguration can be accommodated locally at the modeling level. Since the global generalized inertia matrix is assembled from body-level contributions expressed through the corresponding Jacobians and spatial inertia matrices, modifications in geometric or inertial properties of a body do not require re-deriving the full model from the beginning. Instead, the affected local terms can be updated and reassembled into the global dynamics. In this sense, the framework is particularly suitable for reconfigurable multibody systems in which the system architecture is preserved while component-level parameters or dimensions are altered.

From the control perspective, the modularly assembled equations of motion were used to design a robust model-based controller for trajectory tracking. Although the controller is written in a global form, the framework supports systematic controller synthesis through a model whose terms remain traceable to local body-level contributions. The simulation results on the series-parallel manipulator, under uncertainty and external disturbance, demonstrated the expected boundedness and tracking behavior, supporting the applicability of the proposed framework for trajectory-tracking control of reconfigurable multibody systems with closed kinematic chains.

The main novelty of the proposed approach therefore lies in combining minimal-coordinate modular assembly based on body-level inertial contributions, avoidance of explicit constraint-force calculation when the auxiliary-to-minimal mapping is available, and local updatability of the model under reconfiguration of geometric or inertial body properties.


\section*{Acknowledgement}

We acknowledge the financial support of the Finnish Ministry of Education and Culture through the Intelligent Work Machines Doctoral Education Pilot Program (IWM VN/3137/2024-OKM-4).

\bibliography{ref.bib}
\bibliographystyle{ieeetr}

\end{document}